\documentclass[twoside,twocolumn,9pt]{article}
\usepackage{extsizes}
\usepackage[super,sort&compress,comma]{natbib} 
\usepackage[version=3]{mhchem}
\usepackage[left=1.5cm, right=1.5cm, top=1.785cm, bottom=2.0cm]{geometry}
\usepackage{balance}
\usepackage{mathptmx}
\usepackage{sectsty}
\usepackage{graphicx} 
\usepackage{lastpage}
\usepackage[justification=justified,singlelinecheck=false,font={stretch=1.125,small,sf},labelfont=bf,labelsep=space]{caption}
\usepackage{float}
\usepackage{fancyhdr}
\usepackage{fnpos}
\usepackage[english]{babel}
\addto{\captionsenglish}{%
  
}
\usepackage{array}
\usepackage{droidsans}
\usepackage{charter}
\usepackage[T1]{fontenc}
\usepackage[dvipsnames]{xcolor}
\usepackage{setspace}
\usepackage[compact]{titlesec}
\usepackage{hyperref}

\newcommand{\dz}{\ensuremath{\Delta z_\text{sym}} }

\usepackage{epstopdf}

\definecolor{cream}{RGB}{222,217,201}

\usepackage{siunitx}
\begin{document}

\pagestyle{fancy}
\thispagestyle{plain}
\fancypagestyle{plain}{
\renewcommand{\headrulewidth}{0pt}
}

\makeFNbottom
\makeatletter
\renewcommand\LARGE{\@setfontsize\LARGE{15pt}{17}}
\renewcommand\Large{\@setfontsize\Large{12pt}{14}}
\renewcommand\large{\@setfontsize\large{10pt}{12}}
\renewcommand\footnotesize{\@setfontsize\footnotesize{7pt}{10}}
\makeatother

\renewcommand{\thefootnote}{\fnsymbol{footnote}}
\renewcommand\footnoterule{\vspace*{1pt}%
\color{cream}\hrule width 3.5in height 0.4pt \color{black}\vspace*{5pt}} 
\setcounter{secnumdepth}{5}

\makeatletter 
\renewcommand\@biblabel[1]{#1}            
\renewcommand\@makefntext[1]%
{\noindent\makebox[0pt][r]{\@thefnmark\,}#1}
\makeatother 
\renewcommand{\figurename}{\small{Fig.}~}
\sectionfont{\sffamily\Large}
\subsectionfont{\normalsize}
\subsubsectionfont{\bf}
\setstretch{1.125} 
\setlength{\skip\footins}{0.8cm}
\setlength{\footnotesep}{0.25cm}
\setlength{\jot}{10pt}
\titlespacing*{\section}{0pt}{4pt}{4pt}
\titlespacing*{\subsection}{0pt}{15pt}{1pt}

\fancyfoot{}
\fancyfoot[LE]{\footnotesize{\sffamily{\thepage~\textbar\hspace{3.45cm} 1--\pageref{LastPage}}}}
\fancyhead{}
\renewcommand{\headrulewidth}{0pt} 
\renewcommand{\footrulewidth}{0pt}
\setlength{\arrayrulewidth}{1pt}
\makeatletter 
\newlength{\figrulesep} 
\setlength{\figrulesep}{0.5\textfloatsep} 

\newcommand{\topfigrule}{\vspace*{-1pt}%
\noindent{\color{cream}\rule[-\figrulesep]{\columnwidth}{1.5pt}} }

\newcommand{\botfigrule}{\vspace*{-2pt}%
\noindent{\color{cream}\rule[\figrulesep]{\columnwidth}{1.5pt}} }

\newcommand{\dblfigrule}{\vspace*{-1pt}%
\noindent{\color{cream}\rule[-\figrulesep]{\textwidth}{1.5pt}} }

\makeatother

\twocolumn[
  \begin{@twocolumnfalse}
{
}\par
\vspace{1em}
\sffamily
\begin{tabular}{m{2cm} m{14cm} m{2cm}}
 & \centering\noindent\LARGE{\textbf{Designing single-layer PDMS devices for micron to millimeter-scale deformations$^\dag$}} \\
\vspace{0.5cm} & \vspace{0.5cm} \\
\vspace{0.5cm} & \vspace{0.5cm} \\

 & \noindent\large{Leon V. Gebhard,\textit{$^{ab}$} Alexandre S. Avaro,\textit{$^{ab}$} Gabriel Amselem,\textit{$^{a}$} and Charles N. Baroud\textit{$^{ab}$}}$^{\ast}$ \\ 
\vspace{0.1cm} & \vspace{0.1cm} \\


 & \noindent\centering\normalsize{
The elasticity of PDMS has played a central role in advancing important microfluidic technologies,
ranging from early valves to sophisticated organ-on-a-chip systems. However, most deformable
microfluidic devices are based on geometries that require complex multi-layer PDMS architectures and
include thin membranes, leading to difficult microfabrication and poor stability. Recently, Jain, Belkadi et al. (Biofabrication 16.3 (2024): 035010) introduced a single-layer PDMS device in which a wide and long microfluidic channel was deformed by pressurizing two adjacent air chambers. While they
demonstrated how the channel ceiling deformation can be leveraged to compress biological materials, it remains unknown how the device geometry influences this deformation. Here, a systematic numerical study is
performed on 14,336 variants of this device, through which the height of the PDMS layer is identified
as the main feature that determines the ceiling deformation. Three modes of channel deformation are
identified as the geometry are varied: a U shape with a central minimum, a W shape with two minima
and a central maximum, or an inverse U shape with an upward-bulging single maximum. The
numerical results are validated in experiments that reproduce the three modes for the predicted
geometries and demonstrate vertical ceiling deformations ranging from a few microns to the millimeter scale. The generality of this approach is demonstrated for two example applications: A fully closing single-layer microfluidic valve and an optical lens of controllable anisotropic magnification. This work leverages the rapid prototyping enabled by 3D printing or micro-milling to open new perspectives in microfluidic actuation.} \\

\end{tabular}

 \end{@twocolumnfalse} \vspace{0.6cm}

]

\renewcommand*\rmdefault{bch}\normalfont\upshape
\vspace{-1cm}

\footnotetext{\textit{$^{a}$~Laboratoire d'Hydrodynamique (LadHyX), CNRS, Ecole Polytechnique, Institut Polytechnique de Paris, 91120 Palaiseau, France.}}
\footnotetext{\textit{$^{b}$~Institut Pasteur, Université Paris Cité, Physical Microfluidics and Bioengineering, 25-28 Rue du Dr Roux, 75015 Paris, France. }}

\footnotetext{\dag~Supplementary Information available: [details of any supplementary information available should be included here]. See DOI: 00.0000/00000000.}
\footnotetext{$\ast$ charles.baroud@polytechnique.edu}





\section*{Introduction}

Many microfluidic technologies have been developed to mechanically deform the channels, starting with the production of externally-controlled valves and pumps,~\cite{Unger2000,Grover2003,Laser2004,utharala2022microfluidic} and followed more recently by devices that allow the confinement,~\cite{liu_confinement_2015} compression, \cite{CompressionGradients, SemiconfinedCompression, UnconfinedComp} or stimulation of cellular monolayers~\cite{LungOnChip} and cell-laden hydrogels~\cite{marsano2016beating,paggi_emulating_2022,paggi_joint--chip_2022}. In many of these cases, an external pressure source is used to deform an elastic material along predetermined directions, in order to achieve channel closure or cell stimulation. As a result the device performance is determined by the elastic properties of the material combined with the geometry and fabrication protocols, with most devices relying on the use of thin membranes and multilayer PDMS architectures. The complexity that is inherent with these designs increases the probability of failure during the microfabrication, which in turn increases the cost and time required to obtain functioning devices, for both academic laboratories and commercial companies. Nevertheless, the ability to deform microfluidic devices on the microscale is  still desirable, namely for mechanobiology applications on individual cells~\cite{roux_equibiaxial_2025} or cell collectives~\cite{grassart2019bioengineered}. In this context, there remains a need for new devices that allow large and well-controlled deformations of microfluidic devices without the need for complex microfabrication.

\begin{figure}[ht!] 
  \centering \includegraphics[width=.5\textwidth]{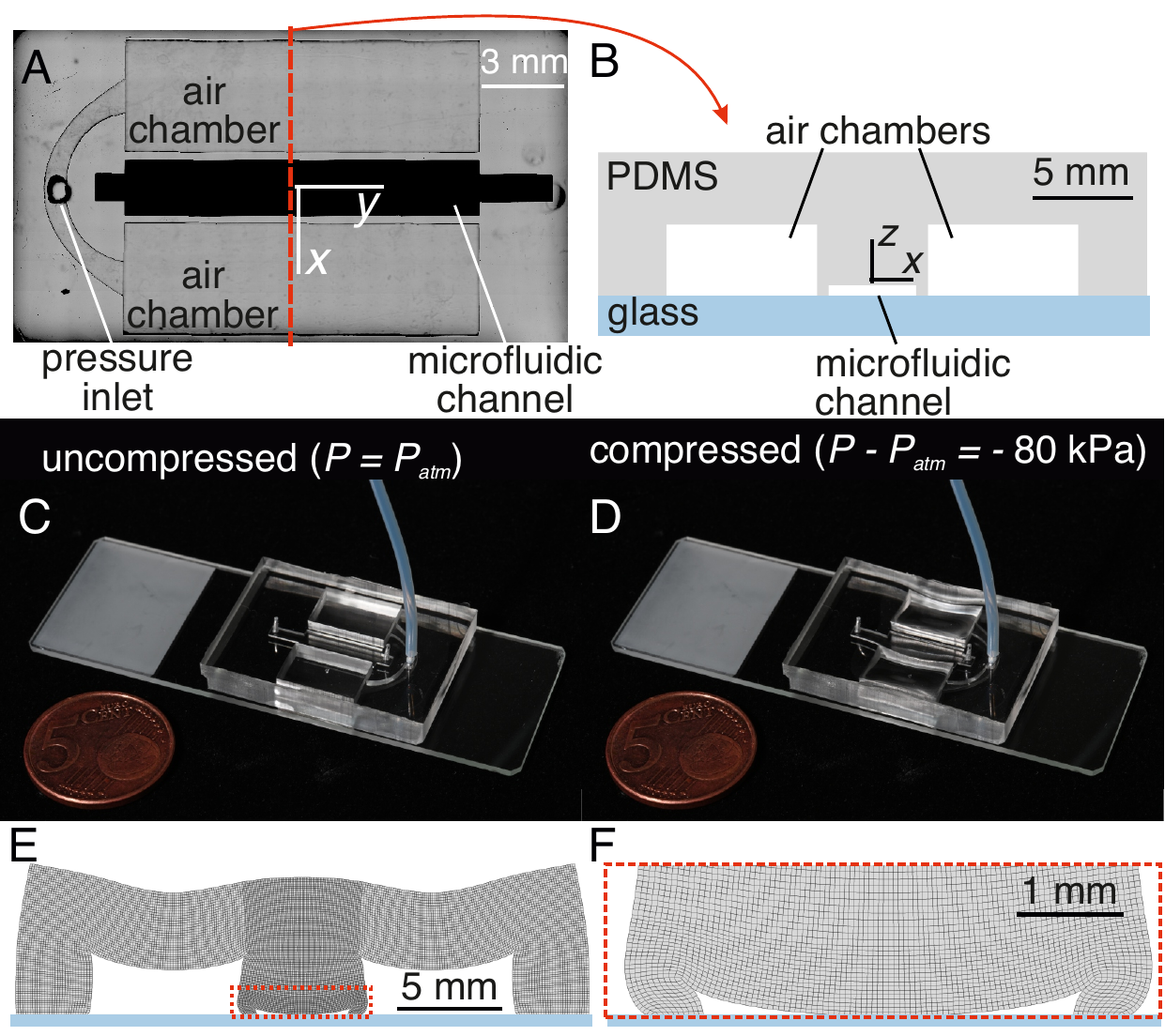}
  \caption{\textbf{Deformable microfluidic chip in the relaxed and deformed state.}
    \textbf{A:} Microscopy image showing the chip geometry.
    \textbf{B,E,F:} Cross section of the chip in relaxed (B) and deformed (E) states as predicted by FEM simulations, with a close up of the deformed channel (F).
    \textbf{C,D:} Images of the chip without (C) or with (D) negative pressure in the air chambers, with a five-cent euro coin for scale.
    }
        \label{fig:Intro}
\end{figure}

One such device was presented recently by Jain, Belkadi et al.,~\cite{MainPaper} who demonstrated a microfluidic channel with a deformable ceiling that was suitable for the compression of cell spheroids, and a similar geometry was later introduced to manipulate droplets and microgels~\cite{saita2025deformation}. Importantly, these devices were produced in a single molding step by pouring PDMS into a 3D-printed mold making it robust and easy to produce. 
Both studies~\cite{MainPaper, saita2025deformation} however relied on a single geometry and did not question the design choices beyond showing their suitability to the task at hand. It therefore remains unclear how general the demonstrated behavior is, and what other geometric choices could be used to modify the device performance.


Here, we explore the role of different geometric parameters on the shape and amplitude of the microfluidic channel's deformation. This exploration is performed through systematic geometric variations of a base chip design and observing how different relevant dimensions influence the deformation of the PDMS structure. By scanning the parameters over a wide range of numerical values, we discover new deformation modes that were not reported previously. The numerical results are then validated experimentally by fabricating devices with different geometries and measuring their respective channel deformations. Finally, two applications of these deformable micro-devices are demonstrated: A single-layer valve and an anisotropic optical lens.


\section*{Exploring deformation regimes}

\subsection*{Device geometry and description of the numerical model}

The base geometry investigated in the current study is based on a simplified version of the device of Jain, Belkadi et al. ~\cite{MainPaper}. It consists of a long, shallow microfluidic channel flanked by two high air chambers of equal length, as shown  Fig. \ref{fig:Intro}A and B. The device's pneumatic circuit consists of two air chambers connected to a single pressure controller, and is fluidically isolated from the microchannel. Correspondingly, a negative air chamber pressure induces a deformation of the entire monolithic device (Fig. \ref{fig:Intro}C and D) but does not directly set the pressure inside the microfluidic channel. 

Given the large size contrast between the streamwise $y$ dimension, compared with the spanwise width $x$ and vertical depth $z$ (see Fig.~\ref{fig:Intro}A,B), we model the device's cross-sectional deformation  under a plane-strain assumption. The resulting two-dimensional model is simulated using a finite element method in Abaqus (Dassault Systemes, France) as shown in Fig. \ref{fig:Intro}E,F. Note that it is sufficient to simulate only half of the domain, because of the device's symmetry with respect to the central $zy$-plane. 

In this model, PDMS is described as an incompressible neo-Hookean material, as this material law describes moderate deformations more accurately than linear elasticity but is still fully defined by a single stiffness paramterer $C$. Simulations were performed with $C=204$~kPa to match the deformations observed experimentally. The Young's modulus equivalent to this value of $C$ is $E=0.6$~MPa\cite{holzapfel}, which is in the lower range of previously reported stiffness for PDMS.\cite{PDMS_Stiffness1,PDMS_Stiffness2,PDMS_Stiffness3,PDMS_Stiffness4} The device's self-contact and its contact with the glass are described as frictionless, and the glass is treated as rigid body. The PDMS-glass bond was captured by fixing both rotation and displacement at the bottom of the device. Detailed descriptions of the meshing, along with the convergence study, are provided in the Material and Methods section.

\subsection*{Geometric features and channel deformation}

\begin{figure*}[h!]
    \centering
    \includegraphics[width=\linewidth]{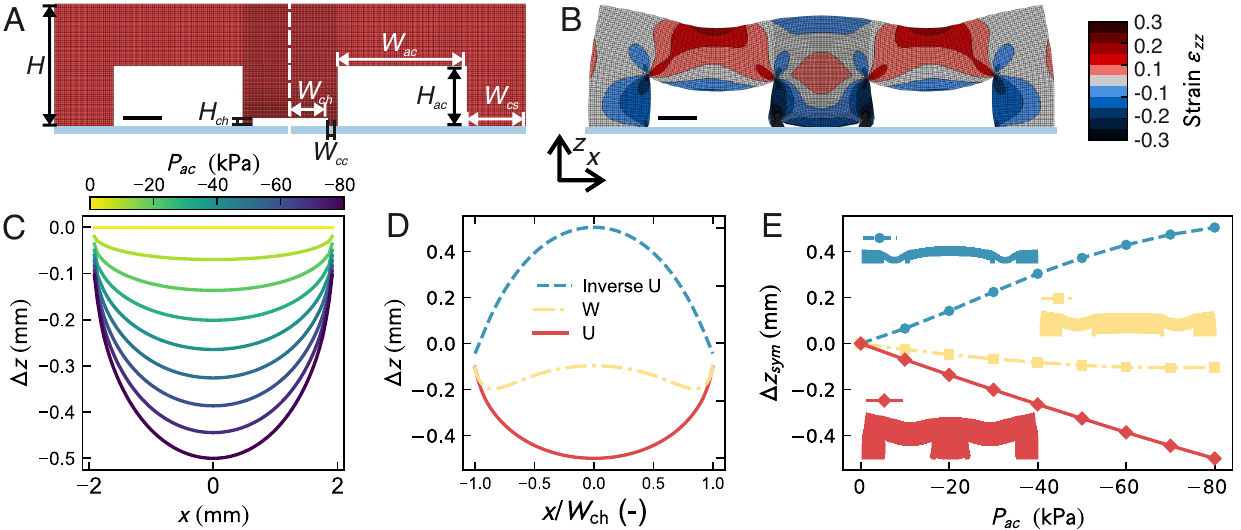}
    \caption{\textbf{The numerical model predicts different deformation modes based on the geometric features.}
    A: Geometry of the 2D model. $H:$ PDMS layer height, $H_{\text{ac}}:$ air chamber height, $H_{\text{ch}}:$ channel height,  $W_{\text{ac}}:$ air chamber width, $W_{\text{ch}}:$ channel half width, $W_{\text{cc}}:$ distance between channel and air chamber, $W_{\text{cs}}:$ distance between air chamber and side wall.
    B: Example result of a FEM simulation. Shown is the deformed mesh color-coded with vertical strain $\varepsilon_{zz}$. Scale bars in A and B correspond to $2$ mm.
    C: Deformation profile of the microfluidic channel ceiling for air chamber pressures  between $0$ and $-80$ kPa. 
    D: Channel ceiling deformation at $-80$ kPa for the three different geometries sketched in E. 
    E: Vertical deformation of the microfluidic channel ceiling at the central symmetry plane $\Delta z_\text{sym} = \Delta z(x=0)$ for three geometries as function of the air chamber pressure.}
    \label{fig:sobol}
\end{figure*}

The microdevice geometry is defined by a large number of geometric parameters, whose variation influences the deformation profile of the microchannel. The current study explores the impact of six of these parameters, as sketched in Fig.~\ref{fig:sobol}A: The total height of the PDMS layer ($H$); the half-width of the microfluidic channel ($W_{\text{ch}}$); the height and width of the air chambers (respectively $H_{\text{ac}}$ and $W_{\text{ac}}$); the distance between the air chambers and the microfluidic channel ($W_{\text{cc}}$), as well as the distance between the air chamber and the side wall of the PDMS ($W_{\text{cs}}$). 

For a given set of values of geometric parameters, the numerical model computes the deformation of the whole microdevice, following the protocol described above. An example deformed device is shown in Fig.~\ref{fig:sobol}B, in which the colors indicate the vertical strain, with red and blue respectively indicating extension and compression. This strain field shows that although the pressure is applied only in the air chamber, the deformations extend over the entire device, leading to a large decrease in the microfluidic channel height. 

The amplitude of channel ceiling deformation depends on the pressure applied to the air chambers. Fig.~\ref{fig:sobol}C shows the displacement of the channel ceiling at different air chamber pressures, for the example geometry of Fig.~\ref{fig:sobol}B. We observe that the deformation magnitude increases with higher pressures but the general shape of the ceiling conserves its $U$ shape, with one minimum in the center, for all values of the pressure. 

However, the $U$ shaped ceiling deformation shown in Fig.~\ref{fig:sobol}B and C is not universal for all device geometries. Indeed, by varying the different parameters, it is possible to observe qualitatively different deformation modes, as shown in  Fig.~\ref{fig:sobol}D and E: The channel ceiling of a thick device ($H/W = 0.525$) with wide air chambers and a narrow channel deforms in the previously described '$U$-shape'. In contrast, a thin ($H/W=0.15$) device with narrow air chambers and a wide channel shows a similar shape in the opposite direction, which we correspondingly term 'inverse $U$-shape'. For a device of moderate thickness ($H/W=0.275$) and channel width, the deformed channel ceiling assumes a '$W$-shape', with two minima near the channel edge and a maximum in the center. 

In all three cases, the amplitude of the ceiling displacement at the channel mid-plane $\Delta z_{\text{sym}}$ scales nearly linearly with the applied pressure, as shown in Fig.~\ref{fig:sobol}E. From this observation we conclude that the mode of the deformation is determined by the geometric parameters of the device, while the air chamber pressure mostly affects the deformation amplitude. In the following, the space of geometric parameters is systematically explored to determine which of these parameters determine the mode of deformation of the device.


\subsection*{Identifying the geometric features whose variations dominate channel deformation}

The impact of different geometric parameters can be estimated by varying each one of them individually and computing the resulting channel deformation (see SI Fig.~\ref{SIFig:Sweep}). However, understanding their relative impact and interactions requires a global sensitivity analysis. Here, given the large number of parameters, we turn to Sobol's method  to identify which parameters have the strongest influence on the deformation of the microfluidic channel.~\cite{Sobol2001-rw} The geometric parameters are used as the inputs of the analysis, while the ceiling deformation at the symmetry plane ($\Delta z_{\text{sym}}$) is chosen as system output. The algorithm then computes the variation in \dz  as the parameters are varied individually or in combination.~\cite{saltelli2008global}  The individual Sobol indices $S_1$ quantify the relative contribution of individual parameters, while the total Sobol indices $S_T$ estimate the importance of a parameter both alone and through interaction with the others. The simulations for the global sensitivity analysis are performed at a driving pressure at $P_{\text{ac}}=-10$~kPa. 

A key advantage of Sobol's method is the gain in computational efficiency. While for a classical grid search the number of simulations needed to evaluate $n$ different values for $k$ parameters scales exponentially ($\mathcal{O}_\text{grid}(n^k)$), Sobol's method varies the input parameters simultaneously leading to a linear scaling of computation time ($\mathcal{O}_\text{Sobol}(n(2+2k))$)~\cite{SobolApplied}. This linear scaling enabled the study of $n=2^{10}$ different values for the $k=6$ geometric parameters in $n(2+2k)=14,336$ simulations, instead of the $n^k = (2^{10})^6 \approx10^{18}$ simulations required for a naïve grid search.

The geometric parameters are varied independently in a range that reflects standard dimensions for microfluidic chips, as listed in Table~\ref{tab:Sobol}. The range for most of the parameters is kept between $0.5$ and $10$~mm. However, three parameters were further constrained: The channel height ($H_{\text{ch}}$) was kept constant at $0.5$~mm, the lower limit of the total PDMS height was set to $H_{\text{ch}}+0.5$~mm, and the air chamber height was defined relative to the total height $H_{\text{ac}}/H$.

\begin{table}
\caption{\label{tab:Sobol}Upper and lower limits of the parameter space explored using the Sobol sensitivity analysis.  }
\begin{tabular}{lccccccc}
 &$W_{\text{ch}}$&$W_{\text{ac}}$&$W_{\text{cc}}$&$W_{\text{cs}}$&$H_{\text{ch}}$&$H$&$H_{\text{ac}}$\\
\hline
Min. (mm)&  $0.5$ & $0.5$ &$0.5$ &$0.5$ & $0.5$ &$H_{\text{ch}}+0.5$ & $0.05H$\\
Max. (mm)&  $10$ & $10$ &$10$ &$10$ & $0.5$ &$H_{\text{ch}}+10$ & $0.95H$\\
\end{tabular}
\
\end{table}
The individual ($S_1$) and total ($S_T$) Sobol indices resulting of the global sensitivity analysis are shown in Fig. \ref{fig:ThreeShapes}A. From the total Sobol indices, it is found that the parameters with the strongest influence on the ceiling deformation are the PDMS layer height $H$, the distance between the microfluidic channel and the air chamber $W_{\text{cc}}$, the air chamber width $W_{\text{ac}}$, and the channel width $W_{\text{ch}}$. In contrast, the width of the outer side wall $W_{\text{cs}}$ and the height of the air chamber $H_{\text{ac}}$ seem to have negligible influence on the deformation of the ceiling.  Correspondingly, for the rest of this study, they will be held constant using arbitrary values convenient for fabrication:  $H_{ac}/H=0.5$ and $W_{cs}=3$ mm. 

\subsection*{Mapping the range of achievable ceiling deformations}
 
\begin{figure*} 
  \centering
  \includegraphics[width=1\textwidth]{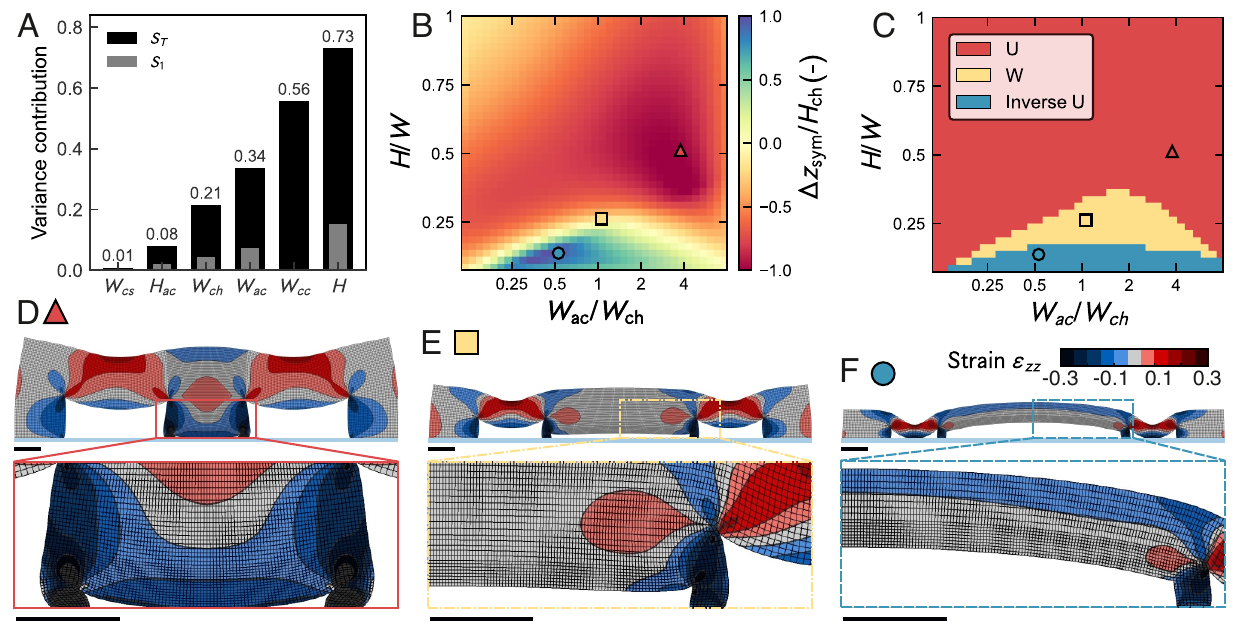}
  \caption{\textbf{Changes in two geometric features lead to qualitatively different deformations.}
  A: Sensitivity of $\Delta z_{\text{sym}}$ towards the geometrical parameters according to a Sobol sensitivity analysis. Shown are individual and total Sobol indices for all geometric parameters varied in this study.
    \textbf{B,C:} Channel ceiling deformation at the symmetry plane $\Delta z_{\text{sym}}$ (B) and deformation mode (C) as function of the non-dimensional PDMS height $H/W$ and the ratio of air chamber to channel width $W_{\text{ac}}/W_{\text{ch}}$ at $P_\text{ac} = -80$ kPa. 
    \textbf{D,E,F:} Deformed chip according to the 2D FEM simulation at $P_\text{ac} = -80$ kPa for geometries whose ceiling deform in a U (D), W (E) or inverted U (F) shape. The colors indicate vertical stretch (red) and compression (blue). The scale bars in panels D-F are of length $2$ mm.}
        \label{fig:ThreeShapes}
\end{figure*}

The results presented above leave us with four dimensional parameters that influence the chip deformation. The dimensionality of this space can be further reduced based on practical experimental considerations. First, we note that larger ceiling deformations are obtained for smaller values of $W_{\text{cc}}$ (see SI Figs~\ref{SIFig:Sweep} and~\ref{fig:WccSI}). Therefore, we set $W_{\text{cc}}=0.5$~mm, a low value that remains practical from a  microfabrication perspective. Second, we introduce the total width of the device by setting the $x$-dimension of the simulation domain to $W=12$~mm, such that the device can fit on a standard microscope slide. This constraint imposes that the sum of the individual widths must sum up to $W$, which leads to the relation $W_{\text{ch}}+W_{\text{ac}}= 8.5$~mm. The remaining free parameters are finally written in dimensionless form as $H/W$ and $W_\text{ac}/W_\text{ch}$.

The numerical model is used to simulate \dz as these dimensionless parameters are varied for a driving pressure   $P_{\text{ac}}=-80$~kPa. The resulting map of $\dz/H_{\text{ch}}$ (Fig.~\ref{fig:ThreeShapes}B) shows a wide variety of deformation amplitudes, ranging from negative displacements, where the microchannel ceiling touches the glass slide, to positive values, where the ceiling bulges upward. 

These deformation amplitudes, in fact, correspond to the three qualitatively different deformation modes, previously shown in Fig.~\ref{fig:sobol}D: the $U$, $W$ and inverse $U$ shaped ceiling deformations. The transition between these three deformation modes is summarized in the phase space Fig. \ref{fig:ThreeShapes}C. 

Further insights into the channel deformation mechanism can be obtained from the global strain field of the deformed structures, as shown in Fig. \ref{fig:ThreeShapes}D-F. These results highlight the alternation between regions of vertical extension (red) and compression (blue) within the PDMS, which in turn induce the deformation of the microchannel ceiling. These strain fields indicate that the distribution of strain varies strongly as the geometry is modified. 




\subsection*{Experimental validation of the numerical results}

\begin{figure}[h!] 
  \centering
  \includegraphics[width=0.5\textwidth]{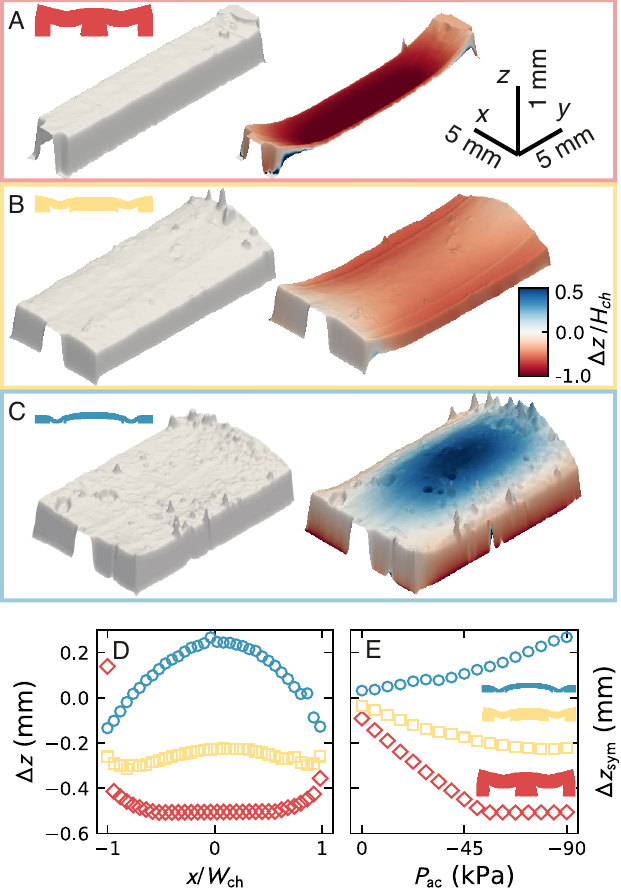}
  \caption{\textbf{Fluorescence intensity measurements confirm the predicted deformation shapes.}
    \textbf{A-C:}  Measurements of the relaxed  and deformed channel geometry, where A-C respectively confirm the predicted $U$ (A), $W$ (B) and inverse $U$ (C) shape. D: Deformation profiles along the $x$ direction in the streamwise ($y$) central $5\%$ of the channel. E: Deformation of the channel ceiling's center point as function of the air chamber pressure $P_\text{ac}$.
    }
        \label{fig:Exp}
\end{figure}

The deformation modes predicted by the numerical model were validated experimentally using three example devices whose cross-sections correspond to those shown in  Fig.~\ref{fig:ThreeShapes}D-F. The devices were fabricated by molding a standard mixture of PDMS (10:1 base:cross-linker, cured at 70$^\circ$C for 2 hours) on 3D-printed masters (Asiga printer, PlasGray v2 resin) and plasma bonding the PDMS layers on glass microscope slides.

The local channel depth was then quantified by \textit{fluorescence intensity profilometry}, as described in the Materials and Methods section. In short, the channels were flushed with a fluorescein solution and the local channel height was then inferred by performing a flat-field correction on fluorescence images. The images were taken with a low numerical aperture objective (2$\times$, NA=0.06) in order to maximize the depth of field. A scan of the complete microfluidic channel was obtained for applied pressures ranging between $0$ and $-90$ kPa.

From this fluorescence intensity profilometry we compute the shape and channel ceiling deformation for three different device geometries. The experimental deformation profiles for the three produced devices respectively reproduce the $U$, $W$ and inverse $U$ modes predicted by the model, as shown by the color maps of Fig.~\ref{fig:Exp}A-C (see also the SI movies).

In the case of the $U$ mode (Fig.~\ref{fig:Exp}A), the experimental data reveal that the downward deformation does not only follow the $U$-shape in the cross sectional, but also along the streamwise direction.  At a high air chamber pressure ($P_{\text{ac}}=-90$~kPa) the channel ceiling contacts the floor. This demonstrates  the large deformations which can be achieved with this design. 


Likewise, the predicted $W$-mode deformation is recovered as the microchannel width is increased and the PDMS layer height is decreased (Fig.~\ref{fig:Exp}B) compared to the $U$-mode chip design. 
The dependence of the ceiling deformation on the streamwise direction is weak except for very local regions near the channel inlet and outlet. 

Finally, the inverse-$U$ mode is observed for a device made from an even thinner layer of PDMS with an even wider channel then the $W$ mode design, as shown in Fig.~\ref{fig:Exp}C. This deformation mode also displays a strong dependence on both the $x$ and $y$ directions, with a maximum near the center of the channel. 

These experimental measurements highlight the importance of the streamwise dimension which was not accounted for by the numerical model. Indeed the edge deformation increases along the streamwise direction from the inlet or outlet to the center of the channel, for all geometries. Nevertheless, the cross-sectional deformation profiles agree with the predicted deformation modes, as shown in Fig.~\ref{fig:Exp}D, where the deformation profiles are computed as a median of the central $5\%$ of the channel's $y$-dimension. Furthermore, the deflection of the microchannel ceiling's central point along the $x$ and $y$-direction recovers the almost linear scaling predicted scaling from Fig. \ref{fig:sobol}D.





\section*{Applications of the deformable devices}

While the sensitivity analysis above provides insights into the most important design parameters, novel functionalities are enabled by further relaxing the imposed geometrical constraints.  Below we introduce two designs where additional geometric parameters are introduced in order to perform new operations.  

\subsection*{Fully closing microfluidic valve}
\begin{figure}[h] 
  \centering
  \includegraphics[width=\textwidth/2]{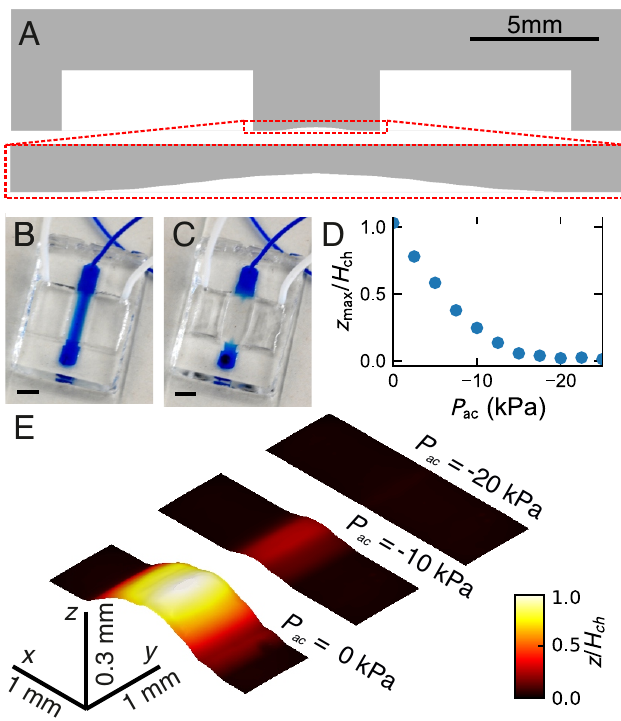}
  \caption{\textbf{Single layer microfluidic valve  with a bell-shaped channel ceiling.}
    \textbf{A:} Cross section of the undeformed valve with an inset showing the channel geometry.
    \textbf{B,C:} Images of the microchannel filled with ink in the opened (B) and closed (C) conditions.
\textbf{D:} Experimental  measurement of the maximal height of the valve cross section as function of the air chamber pressure. 
    \textbf{E:} Valve shape at three different air chamber pressures based on fluorescence intensity  profilometry.}
        \label{fig:Valve}
\end{figure}

While the channel ceiling can be brought into contact with the bottom glass slide, the sharp edges of the channel prevent full closure (see Fig. \ref{fig:ThreeShapes}D and~\ref{fig:Exp}A). Nevertheless, it is possible to completely block the flow inside the device by modifying the shape of the channel ceiling, effectively resulting in a microfluidic valve. Such a device is shown in Fig.~\ref{fig:Valve}, where the flat channel ceiling is replaced by a "bell-shaped" ceiling, i.e., a channel geometry without sharp corners (see Fig.~\ref{fig:Valve}A). 

This channel fully closes when sufficient negative pressure is applied in the air chambers. This is first shown in Fig.~\ref{fig:Valve}B and C, in which blue ink is initially  (B) injected in the microchannel and then  fully ejected upon actuation of the chip (C). Fluorescence intensity profilometry further provides a quantitative measure of the progressive closing for different values of the pressure (Fig.~\ref{fig:Valve}D, E). Indeed, the low pressure required to close the valve implies that it  not only functions as  variable hydraulic resistance but can also serve as an on-off valve.


\subsection*{Deformable channels as tuneable anisotropic lenses}

In all of the above discussion, the cross-section was assumed to represent a channel that extends linearly in the third dimension, with a geometry inspired by microfluidics. This assumption of a linear geometry does not need to be maintained, as the cross-section can also represent a system with rotational symmetry. A miniature deformable device with a circular geometry is well suited as an optical element, which may have many potential applications in microscopy or in consumer optical devices, such as cell phones or other cameras.~\cite{hampson_adaptive_2021,zhang_adaptive_2023}

\begin{figure}[h] 
  \centering
  \includegraphics[width=\textwidth/2]{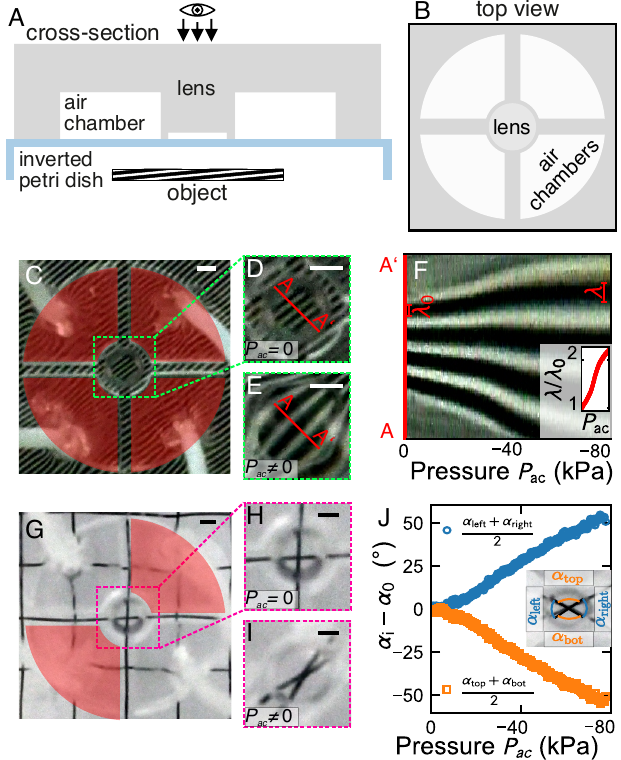}
  \caption{\textbf{Pneumatically tunable optical element.}
    \textbf{A,B:} Cross-section (A) and top view sketches (B) of the PDMS optical device. The device consists of a circular lens region, surrounded by an annular set of air chambers.
    \textbf{C,D,E:} Images of the relaxed device above a striped object in the relaxed $P_{ac} = 0$ (C,D) and actuated state $P_{ac} = -80$ kPa (E).
    \textbf{F}: Kymograph of the highlighted A-A' line over pressure. The inset shows the magnification as a function of the driving pressure.  In C and G, only the air chambers highlighted in red correspond are actuated.
\textbf{G,H,I:}  Lens above a grid pattern in the relaxed (G,H) and anisotropically actuated (I) states. 
\textbf{J:} Quantification of the image distortion. All unlabeled scale bars correspond to $2$ mm. 
}
        \label{fig:Lense}
\end{figure}

Such a lens with tunable optical properties is shown in Fig.~\ref{fig:Lense}. The cross-section of this device conserves the general pattern of the previous devices (Fig.~\ref{fig:Lense}A) but the top view now consists of a central circular "channel", surrounded by an annular air chamber partitioned in four sections (Fig. \ref{fig:Lense}B). In the relaxed state of ambient air chamber pressure, the PDMS/air interfaces are all flat and allow light to pass through unhindered. When a negative pressure is applied to the air chambers, the chip deformation leads to a curvature of the flat surfaces, which deviates the light passing through the chip. The resulting effect is a tunable lens whose properties can be controlled by an external pressure. 

To demonstrate this effect, the device is placed on an inverted Petri dish above a regularly striped pattern, as shown in Fig. \ref{fig:Lense}A and C. The Petri dish serves as a spacer between the lens and the object. When negative pressure is applied to all four independent air chambers, the image of the stripes is magnified (Fig. \ref{fig:Lense}D and E). The magnification amplitude is quantified by measuring the width $\lambda$ of the stripe spacing as a function of pressure, as shown in Fig. \ref{fig:Lense}F. In the current device, we estimate that the magnification increases from 1$\times$ to about 2$\times$ as the driving pressure varies from $P_{ac}=0$ to $-80$~kPa, as highlighted in the inset of Fig. \ref{fig:Lense}F.

More interestingly, setting different pressures in the four air chambers leads to an anisotropic deformation of the image. We show in Fig.~\ref{fig:Lense}G-I an example experiment where negative pressure is only applied to two opposing sections while imaging a rectangular grid. This differential forcing leads to a asymmetric deformation, i.e., where the value of the angles is modified, as shown in Fig.~\ref{fig:Lense}H-I. Here again the image distortion can be tuned by fixing the value of the imposed pressure (Fig.~\ref{fig:Lense}J). These measurements show that the  presented device can achieve a strong, tunable, anisotropic image distortion.

\section*{Summary and discussion}

In response to the persistent need for deformable micro-devices, shape control of a microfluidic channel without multilevel fabrication represents an important asset. The current study shows how to achieve deformations that range from microns to millimeters, using simple microfabrication protocols that leverage the opportunities offered by 3D printing or micro-milling techniques for rapid prototyping. As a result, design iterations are fast and inexpensive, therefore allowing many geometric parameters to be explored to tune the deformation mode. Once fabricated, the operation of these devices is also simple, as it only requires the air chambers to be connected to a negative pressure source. It is worth noting that the use of a negative pressure, in contrast with inflating the air chambers, reduces the vulnerability to leaks and bonding failures, as the negative pressure pushes the channel walls onto the glass slide. 

The global sensitivity study on the geometric parameters shows that the channels deform either downwards or upwards with a $U$-shaped ceiling, or have a nearly flat $W$ deformation profile. In our exploration of parameters the mode of deformation is almost exclusively determined by the geometry of the device, independently from the material stiffness and the driving pressure. Interestingly, the thickness of the PDMS layer plays a determining role in selecting the deformation mode, indicating that device height is a critical fabrication parameter. In turn the physical parameters, such as the Young's modulus or the driving pressure, determine the amplitude of the deformation. While this behavior may not be a general feature for all geometries, it was found to be very robust in our devices both numerically and experimentally.

Finally, while the systematic numerical study explored parameter changes around simplified geometry, the two applications show how other geometric modifications can be introduced to perform specific tasks: either in the shape of the microchannel ceiling or in the global shape of the device. The devices of Figs.~\ref{fig:Valve} and~\ref{fig:Lense} are presented here to demonstrate the proof-of-concept, but each of them can now be optimized, e.g., by further modifying the geometry or material properties. More generally, other designs can be conceived to produce other fluidic operations, such as pumping or mixing, or to couple with biological materials for organ-on-a-chip or other mechanobiology applications.


\section*{Materials and Methods}

\subsection*{Numerical Simulations}

Finite element analysis was performed using Abaqus (Dassault Systèmes, France) and automated via its Abaqus-Python interface. The cross-sectional deformation (Fig. \ref{fig:Intro}B,E,F) was approximated using a 2D plane strain model, motivated by the fact that the chips depth significantly exceeds its width and height. 

The PDMS was described as incompressible neo-Hookean solid using $C = 206$~kPa.\cite{holzapfel} This hyperelastic material law was chosen to account for deformations that exceed the small-strain limits of linear elasticity. While more complex models such as the five-parameter Mooney-Rivlin model can capture PDMS behavior with higher fidelity at extreme strains, \cite{PDMS_Models} we chose the neo-Hookean model to maintain the physical interpretability of a single material constant.

The pressure was applied as a 'pressure load' on the internal surface of the air chamber and the bottom of the PDMS was fixed using the 'Encastre' boundary condition. The glass slide was modeled as an analytical rigid surface also fixed in place using the 'Encastre' condition. Both the contact between the glass and PDMS and the self-contact of the PDMS part were defined as hard and friction-free.

To enable the creation of a structured mesh, the geometry was partitioned into rectangular sub-domains. The partitioning strategy also allowed for local mesh refinement in regions where high strain was anticipated or where high resolution of deformation data was desired. This was for instance the case at the channel ceiling and in the region between the air chamber and the microchannel. This partitioning also enabled the design of a mesh composed entirely of quadrilateral elements, whose edges aligned to the principal axes of the coordinate system (see Fig. \ref{fig:sobol}A). The mesh was chosen to be composed of first order reduced hybrid integration (CPE4RH) elements, where the choice of reduced hybrid integration elements is necessary to prevent volumetric locking, as we assume the material to be incompressible.\cite{belytschko2014nonlinear}

The mesh density was evaluated through a convergence study (SI Fig. \ref{fig:convergenceStudy}), in which the element size in the refined regions was varied from 0.025~mm to 0.15~mm. Convergence was assessed by calculating the $\mathcal{L}^1$-norm of the difference in vertical channel ceiling displacement between the finest discretizations and the evaluated mesh sizes. Based on this study, we selected a mesh size of 0.1~mm for regions close to the channel and 0.2~mm for the bulk material to balance numerical accuracy with computational efficiency.

\subsection*{Device Fabrication}\label{subs:DeviceFab}

The devices were manufactured using standard PDMS fabrication techniques. Masters were produced either via digital light processing (DLP) 3D printing (Asiga MAX, PlasGRAY resin, Australia) except for the valve, whose aluminium mold was milled with a CNC (Datron M8Cube, Germany). 3D-printed molds were post-cured under UV light for 5 minutes and baked overnight at 70$^\circ$C to prevent PDMS curing inhibition caused by unpolymerized resin residues. \cite{CuringInhibition} To ensure clean detachment of the PDMS, these molds were functionalized by filling them with Novec 1720 (Sigma-Aldrich, MO, USA), followed by a 30-minute curing at 70$^\circ$C. Milled molds required no post-processing beyond an isopropanol wash.


The PDMS (Sylgard 184, Dow, CA, USA) was prepared at a 10:1 base-to-cross linker ratio, degassed in vacuum for 30 minutes and cured for two hours at 70$^\circ$C. An exception was made for the valve and lens, for which a 20:1 ratio was used to lower stiffness. After curing, $1$~mm diameter holes were punched for the microfluidic channel inlets and outlets, and a $2$~mm diameter hole was created for the air pressure control. Finally, the PDMS slabs and glass microscopy slides were plasma-activated in a vacuum plasma cleaner for one minute and bonded together. The chips were finally post-baked for thirty minutes at 80$^\circ$C.

\subsection*{Device Operation}

The device features a central microfluidic channel flanked by two interconnected air chambers as shown in Fig. \ref{fig:Intro}. The device is actuated by controlling the air chamber pressure using a pressure controller (OB1, Elveflow, France). The actuation leads to a deformation of the entire device, including the microchannel, as shown in Fig. \ref{fig:Intro}D, E and F.

\subsection*{Fluorescence intensity profilometry}\label{subs:DeviceMeasurement}

To experimentally measure the channel deformation a saturated aqueous fluorescein solution (fluorescein sodium salt) was filtered, diluted at a ratio 1:1000 in isopropanol (IPA), and injected into the microfluidic channel. Fluorescence at $\lambda_{\text{em}} = 510$~nm was measured with a ORCA 4.0 Flash camera (Hamamatsu, Japan) for varying air chamber pressures, using a Sola Light Engine (Lumencor, OR, USA,  $3.5$ W maximum power) as excitation light source at  $\lambda_{\text{ex}}=460$~nm. The channel was then flushed with pure IPA and measured again to obtain a dark-field image.


A flat-field correction was applied by using the image with pure IPA as the dark-field image and an image of the undeformed channel filled with the fluorescein solution as the flat-field image.\cite{model2014intensity} The corrected signal was obtain using the following formula:

\begin{equation}
    C_{p}=\frac{R-D_{p}}{F-D_{0}},
\end{equation}
where $C_p$ is the corrected image at a given pressure $p$, $R$ is the acquired image, $D_i$ is the dark-field image (either at pressure $p$ where $i=p$, or in the undeformed configuration $i=0$), and $F$ is the flat-field image.

\subsection*{Tunable Lens quantification}

The magnification of the tunable lens was quantified by placing the device on top of an inverted Petri dish, which enables to set the distance between the device and the image to about $1$ cm. We used a pattern of strips of equal width as an example image. A sequence of images was acquired (Nikon Z fc with Irix 150mm macro lens) during which the air chamber pressure was linearly increased from $0$ to $-90$ kPa. A kymograph orthogonal to the stripes was then obtained using the Fiji\cite{Fiji} 'reslice' function. 

The reslice resulted in a square-wave signal corresponding to the succession of white and black stripes. The wavelength of the square wave increased with the pressure magnitude. A fast Fourier transform was computed on a central Hanning window of the image.\cite{HanningWindow} The inverse of the dominant frequency of the  Fourier transform  corresponds to the principal wavelength, which is proportional to the observed thickness of the stripes. Image magnification is then defined as the ratio of wavelength at pressure $\lambda_\text{P}$ over the initial wavelength $\lambda_0$.

Similarly, the anisotropic image distortion was measured by placing the optical device on an inverted Petri-dish. However, the example image used for this experiment is a rectangular grid, whose center is aligned with the device's channel center. An image sequence was acquired using the same optical setup as the lens characterization above, while the pressure was increased in only two opposing air chambers.  For each image, the grid lines are segmented through an intensity threshold computed using Otsu’s method.\cite{OtsusMethod} Then, the angular position  of the grid extremities is computed, using the lens as center point.  The angles obtained for the first image are subtracted from all other angles to compute the distortion $\alpha_\text{i} - \alpha_0$. The top and bottom angles and left and right angles are summed to distinguish the anisotropy of the deformation: $\alpha_\text{left,right}= \frac{\alpha_\text{left}+\alpha_\text{right}}{2}$, $\alpha_\text{top,bot}= \frac{\alpha_\text{top}+\alpha_\text{bot}}{2}$.

\section*{Author contributions}
C. N. B. and G. A. contributed through conceptualization, writing, funding acquisition, project administration and supervision. A. S. A. and  contributed through conceptualization, methodology, visualization, writing and investigation. L. V. G. contributed through conceptualization, methodology,  software, data curation,  visualization, investigation and writing. 

\section*{Conflicts of interest}

CNB is inventor on two patents related to this work. Other authors declare no conflict of interest.

 \section*{Data availability}

Data beyond the supplementary information is available upon request to the authors. The CAD files of the molds used to fabricate the presented devices are available in the public github repository \href{https://github.com/BaroudLab/Deformable-Chip-Designs}{https://github.com/BaroudLab/Deformable-Chip-Designs}. 

\section*{Acknowledgements}

We thank Hiba Belkadi for her support on the experimental validation and Erik Maikranz for the rich discussions on theoretical modeling of the system. We thank Étienne Jambon-Puillet for support on photography. Furthermore, we appreciate the support of Shunsuke Saita (University of Kyoto) and Andrea Rocchi (Drahi-X FabLab) for advice on chip fabrication. This work was  funded by the European Union (ERC-2023-ADG grant number 101142018 MELCART) and Institut Polytechnique de Paris (PhD Track Bioengineering).



\balance


\bibliography{Bibliography,some_references_charles} 
\bibliographystyle{rsc} 

\clearpage
\newpage

\setcounter{figure}{0}
\setcounter{table}{0}
\makeatletter 
\renewcommand{\thefigure}{S\@arabic\c@figure}
\renewcommand{\thetable}{S\@arabic\c@table}
\makeatother

\section{Supplementary Materials}
\begin{figure}[h!]
    \centering
    \includegraphics[width=\linewidth]{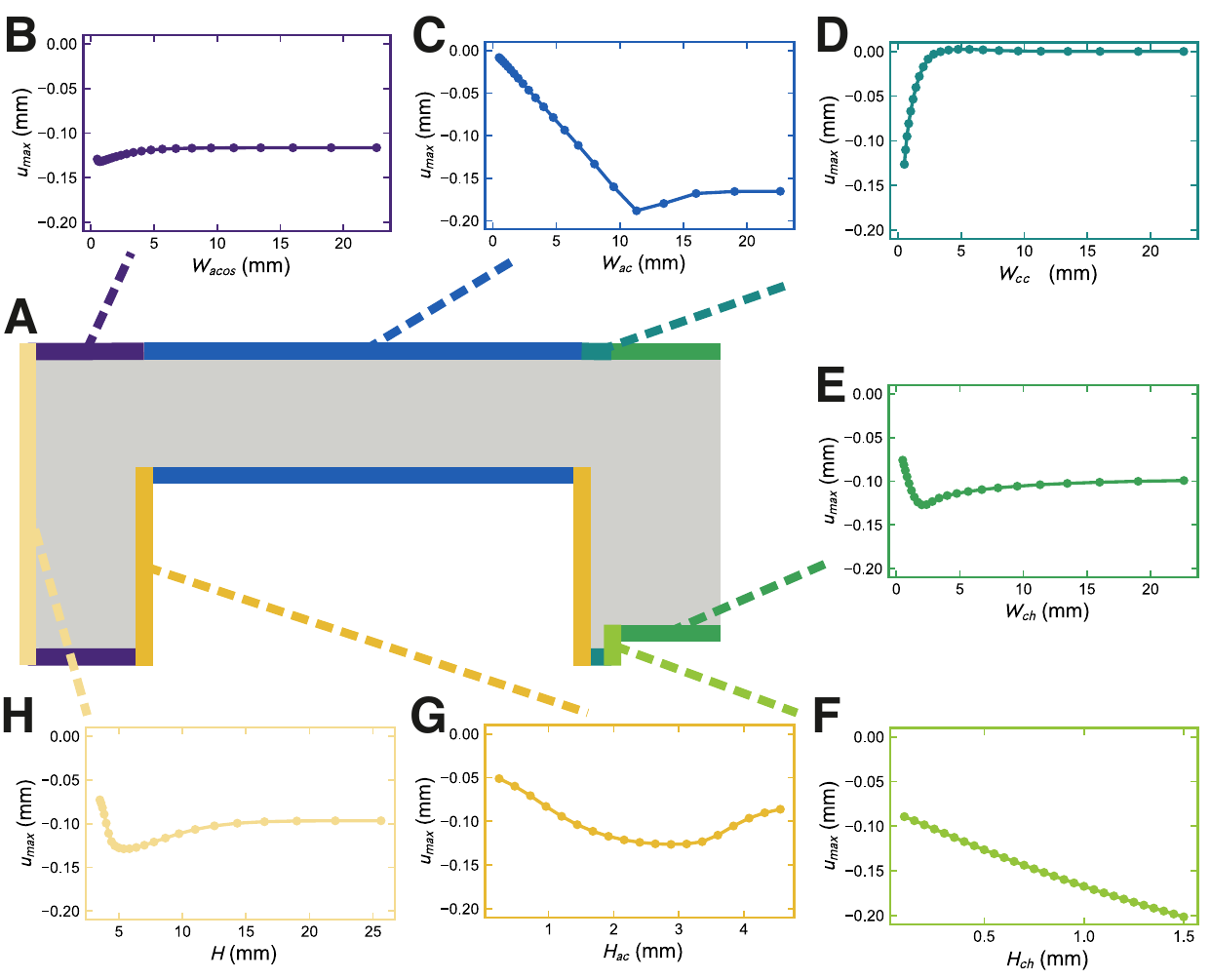}
    \caption{Exploratory simulations reveal the influence of various parameters. A: Half cross section of the device with colored edges indicating the geometrical parameters. B-H: change in maximal downward displacement as function of the indicated parameter, while all others remain constant. }
    \label{SIFig:Sweep}
\end{figure}

\clearpage

\begin{figure}[ht!]
    \centering
    \includegraphics[width=\linewidth]{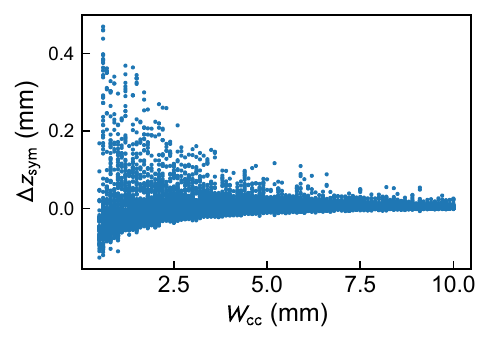}
    \caption{Stochastic distribution of the deformation at the symmetry plane as function of the channel-air chamber distance based on 14,336 simulations with random input geometries.  }
    \label{fig:WccSI}
\end{figure}

\begin{figure}[hb!]
    \centering
        \includegraphics[width=\linewidth]{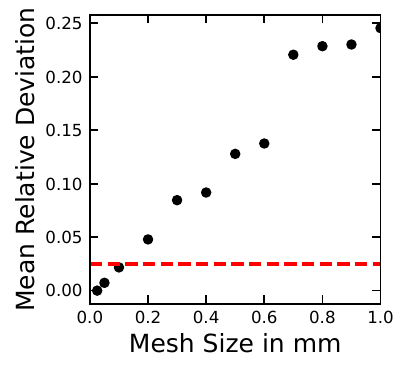}
    \caption{Difference in vertical channel deformation for mesh sizes from $0.25$ mm  and $1$mm, where the deformation obtained using the smallest mesh size is regarded as ground truth. }
    \label{fig:convergenceStudy}
\end{figure}

\begin{figure*}[h] 
  \centering
  \includegraphics[width=1\textwidth]{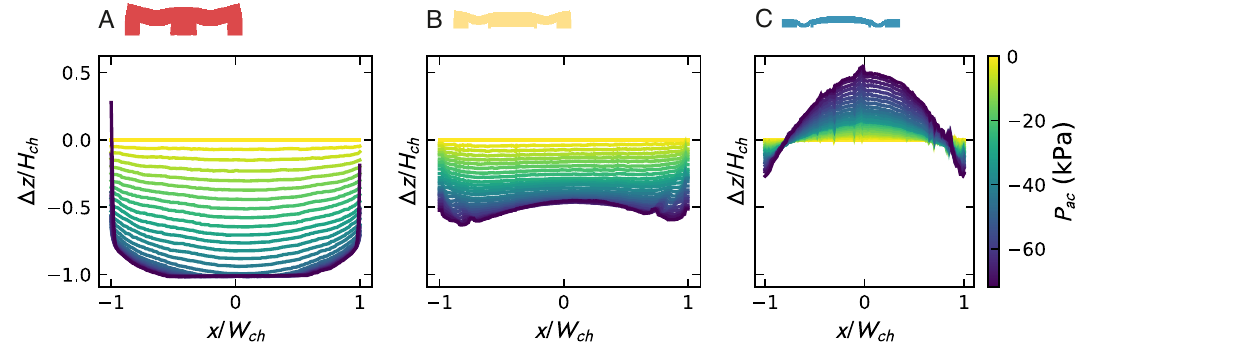}
  \caption{\textbf{Experimentally determined deformation profiles at different driving pressures. (A) $U$-shape, (B) $W$-shape, (C) inverse $U$-shape} }
        \label{SI:sdf}
\end{figure*}

\end{document}